\documentclass{cpcauth}

\usepackage{epsfig}

\usepackage{amssymb}

\newcommand{\dir}{Figs}
\newcommand{\fig}[3]
{
     \noindent
     \unitlength=1mm
     \begin{picture}(#2,#3)
     \put(0,0){\leavevmode \epsfxsize=#2mm \epsffile{\dir/#1}}
     \end{picture}
   \noindent
}

\newcommand{\rr}{ {\bf r} }
\newcommand{\ru}{ {\bf \hat{r}} }
\newcommand{\rb}{ {\bf \hat{b}} }

\newcommand{\uu}{ {\bf u} }

\newcommand{\CCphbulk}
{
\caption{
Bulk phase diagram in pressure-temperature space, calculated from
simulations with three different system sizes as indicated.
}
\label{fig:phbulk}
\bigskip
}

\newcommand{\CCbare}
{
\caption{
Solvent in contact with the bare substrate;
configuration snapshot (top) and number density profile
$\rho$ vs. $z$ (bottom).
}
\label{fig:bare}
\bigskip
}

\newcommand{\CCsnapns}
{
\caption{
Configuration snapshot of grafted chains without solvent 
at the grafting density $\Sigma = 0.34/\sigma_0^2$.
}
\label{fig:snap_nosolv}
\bigskip
}

\newcommand{\CCnosolv}
{
\caption{
Grafted chains without solvent:
Average tilt angle~$\langle \theta \rangle$ of 
the head-to-end vector of the chains 
as a function of the grafting density $\Sigma$.
}
\label{fig:nosolv}
\bigskip
}

\newcommand{\CCsnap}
{
\caption{
Configuration snapshot of grafted chains with solvent at grafting 
density $\Sigma = 0.34/\sigma_0^2$.
Top: Just chain monomers (solvent particles are transparent);
Bottom: Chain monomers (light) and solvent particles (dark).
}
\label{fig:snap}
\bigskip
}

\newcommand{\CCtheta}
{
\caption{
Average tilt angle $\langle \theta \rangle$ as a function 
of the grafting density $\Sigma$.
Top: Head-to-tail vector of grafted chains;
Bottom: Single particles.
}
\label{fig:theta}
\bigskip
}

\begin{document}

\begin{frontmatter}



\title{Surface anchoring on liquid crystalline polymer brushes}


\author{Harald Lange$^{\dag, \ddag}$ and Friederike Schmid$^{\dag,*}$}

\address{
$\dag$ Fakult\"at f\"ur Physik, Universit\"at Bielefeld, 
         33615 Bielefeld, Germany \\
$*$ Telephone: ++49-521-1066191, E-mail: {\tt schmid@physik.uni-bielefeld.de} \\
$\ddag$ Institut f\"ur Physik, Universit\"at Mainz, 
55099 Mainz, Germany
}

\begin{abstract}
We present a Monte Carlo study of the surface anchoring of a nematic fluid 
on swollen layers of grafted liquid crystalline chain molecules.
The liquid crystalline particles are modeled by soft repulsive ellipsoids,
and the chains are made of the same particles. An appropriately modified 
version of the configurational bias Monte Carlo algorithm is introduced, 
which removes and redistributes chain bonds rather than whole monomers. 
With this algorithm, a wide range of grafting densities could be studied.

The substrate is chosen such that it favors a planar orientation
(parallel to the surface). Depending on the grafting density, we find 
three anchoring regimes: planar, tilted, and perpendicular alignment. 
At low grafting densities, the alignment is mainly driven by the substrate.
At high grafting densities, the substrate gradually loses its influence 
and the alignment is determined by the structure of the interface 
between the brush and the pure solvent instead.
\end{abstract}

\end{frontmatter}

\section{Introduction}
\label{sec:introduction}

Liquid crystals have fascinated condensed matter physicists, chemists and 
material scientists for many decades~\cite{degennes,chandrasekhar}
-- first, because they exhibit a large number of beautiful phases with 
intriguing symmetries and interesting material properties, and second, 
because of their technological use, particularly in the domain of display 
devices. One important aspect for technical applications is the 
so-called surface anchoring~\cite{jerome,bahadur,schadt}: Surfaces orient 
nearby molecules, and these in turn orient the whole bulk of the liquid 
crystal. Alignment layers are key components of liquid crystal display 
devices~\cite{schadt}. In particular, one is interested in designing surfaces 
which orient liquid crystals in a well-defined way in any desired direction. 

Possible candidates are liquid crystalline polymer 
brushes~\cite{avi1,avi2,avi3,peng1,peng2}. Halperin and 
Williams~\cite{avi1} have suggested to create a situation 
where a substrate which favors planar anchoring (parallel to the surface) 
competes with a stretched main-chain liquid-crystalline brush which 
favors homeotropic anchoring (perpendicular to the surface).
They predicted that this conflict would result in tilted anchoring 
above a critical grafting density $\Sigma_c$. In their scenario,
the transition between planar and tilted anchoring is continuous. 
Above the transition, the grafting density can be used to tune the 
anchoring angle.

Experimentally, dense liquid crystalline polymer brushes have been
prepared~\cite{peng1}, and their alignment properties have been 
studied~\cite{peng2}. So far, the conditions in the experiments were such 
that the substrate and the brush favor the same type of alignment.
However, a competition as required by the scenario of Halperin 
and Williams can be introduced by a straightforward modification 
of the substrate~\cite{peng2}.

Motivated by that work, we have investigated the consequences
of such a competition by Monte Carlo simulations of an idealized
liquid crystal model. We have studied brushes of liquid crystalline
polymers in a nematic solvent on a substrate which favors planar
anchoring. The relaxation times in such systems are very high. 
Therefore, we have implemented a new version of the configurational 
bias Monte Carlo algorithm~\cite{rosenbluth,frenkel,harald2,chris}: 
Chains are broken up such that their monomers turn into solvent particles, 
and solvent particles are linked together to form new chains.
With this algorithm, a wide range of grafting densities 
could be considered. We shall see that the resulting anchoring scenario
turns out to be quite rich.

We introduce the simulation model in the next section, and describe
the algorithm in the third section. The results are presented and
discussed in section 4. We summarize and conclude in the
last section.
 
\section{The Simulation Model}
\label{sec:model}

Our system consists of soft ellipsoidal particles with elongation
$\kappa = \sigma_{\mbox{\tiny end-end}}/\sigma_{\mbox{\tiny side-side}}=3$.
Two particles $i$ and $j$ with orientations $\uu_i$ and $\uu_j$ separated
by the center-center vector $\rr_{ij}$ interact via the purely
repulsive pair potential 
\begin{equation}
\label{eq:vij}
V_{ij}
= \left\{ \begin{array}{lcr}
4 \epsilon_0 \: (X_{ij}^{12} - X_{ij}^{6}) + \epsilon_0 & : & X_{ij}^6 >
1/2 \\
0 & : & \mbox{otherwise}
\end{array} \right. .
\end{equation}
where $X_{ij} = \sigma_0/(r_{ij}-\sigma_{ij}+\sigma_0)$ and
\begin{eqnarray}
\sigma_{ij}
&=& \sigma_0 \:
\Big\{ \: 1 \: - \frac{\chi}{2}
 \Big[
\frac{(\uu_i\cdot\ru_{ij}+\uu_j\cdot\ru_{ij})^2}
     {1+\chi \uu_i\cdot\uu_j} 
\nonumber \\&& 
\qquad + \:
\frac{(\uu_i\cdot\ru_{ij}-\uu_j\cdot\ru_{ij})^2}
     {1-\chi \uu_i\cdot \uu_j}
\Big] \:  \Big\}^{-1/2}
\end{eqnarray}
with $\chi = (\kappa^2-1)/(\kappa^2+1)$. The function $\sigma_{ij}$ 
approximates the contact distance between the two ellipsoids in the 
direction $\ru_{ij} = \rr_{ij}/r_{ij}$~\cite{berne}.

Solvent and chain particles have the same nonbonded interactions.
In addition, chain monomers are connected by anharmonic springs of length $b$ 
with equilibrium length $b_0$ and a logarithmic cutoff at $|b-b_0|=b_s$. 
At $|b-b_0| < b_s$, the spring potential is given by
\begin{equation}
V_s(b) =  - k_s/2 \:  b_s^2 \ln [ 1 - (b-b_0)^2/b_s^2)\: ].
\end{equation}
The direction of a bond is given by the unit vector $\rb$.
A stiffness potential 
\begin{equation}
V_a (\uu,\rb_1,\rb_2)
= - k_a \:( |\uu\rb_1| + |\uu \rb_2| + 2 \: \rb_1 \rb_2)
\end{equation}
is imposed, which penalizes non-zero angles between the orientation 
$\uu$ of a monomer and its two adjacent bonds $\rb_1$ and $\rb_2$. 
The system is confined by hard walls at two sides $z=0$ and $z=L_z$: 
The distance between the walls and the centers of ellipsoids, which 
have an angle $\theta$ with respect to the surface normal, must exceed
\begin{equation}
d_z(\theta) = \sigma_0/2 \: \sqrt{1+\cos^2(\theta)\: (\kappa^2-1) }.
\end{equation}
The chains are grafted to the walls at one end, the grafting points are
arranged on a regular square lattice.

We have studied systems with roughly $N= 2000$ solvent particles (the 
numbers varied in the different runs) and up to 242 chains of four monomers.
The simulation boxes were rectangular of size 
$L_{\parallel} \times L_{\parallel} \times L_z$ 
with periodic boundary conditions in the lateral directions and fixed
boundary conditions in the $z$ direction. The thickness $L_z$ was allowed 
to fluctuate, so that the simulations could be performed at constant
pressure. The lateral size $L_{\parallel} = 12 \sigma_0$ was kept fixed in 
order to maintain a constant grafting density $\Sigma$. Trial moves included
monomer displacements, monomer rotations, rescaling of $L_z$, and
the configurational bias moves described in the next section.
The initial configuration was set up such that all particles pointed in 
the $x$ direction. The system was then equilibrated over at least 1 million
Monte Carlo steps, and data were collected over 5 or more million
Monte Carlo steps. One ``Monte Carlo step'' includes on average $N$ 
attempts of monomer displacements, $2 N$ attempts of monomer rotations, 
one attempt of $L_z$ rescaling, and one attempt of a configurational bias 
move.

The model parameters chosen were $k_s=k_a=10.$, $b_0=4.$ and $b_s=0.8$.
Here and throughout, we use scaled units defined in terms of
$\epsilon_0$, $\sigma_0$, and the Boltzmann constant $k_B$. 
Unless stated otherwise, the simulations were performed at constant
temperature $T=0.5$ and constant pressure $P=3$. The average number
density in the bulk was \mbox{$\langle \rho \rangle = 0.313$}.

\section{Configurational Bias Monte Carlo in a Solvent}
\label{sec:cbmc}

The most frequently used continuous space version of the configurational 
bias Monte Carlo algorithm~\cite{frenkel,frenkel2} removes chains from the
system and inserts new monomers into the system. However, this method is not 
effective in our system, because the space is densely filled with monomer 
and solvent particles, and little free volume for new particles is available. 
Therefore, we have resorted to a different scheme. Instead of removing the 
monomers of a chain, we only remove the connecting bonds and redistribute them 
such that new chains are formed from solvent particles. This method resembles 
closely the lattice version of the configurational bias Monte Carlo 
algorithm~\cite{frenkel}. The only difference is that the lattice is 
now random, with ``lattice sites'' given by the centers of solvent 
or monomer particles. 

In practice, we proceed as follows: First, we remove all the bonds from
a randomly chosen chain. Then we find all $n_1$ solvent particles, 
which are at a distance $b\in[b_0 - b_s,b_0 + b_s]$ from the grafting 
point, and choose one of them according to the  Boltzmann probability 
\begin{equation}
P_1 = \frac{1}{Z_1}
\exp\Big[- \frac{V_s(b_1) + k_a(1-|\uu_1\rb_1|)}{T}\Big]
\end{equation}
Here $b_1$ is the length of the bond between the grafting point
and the particle, $\rb_1$ its direction, and $\uu_1$ the
orientation of the particle. The normalization factor $Z_1$ is the sum
\begin{equation}
Z_1 = \sum_{j = 1}^{n_1} 
\exp\Big[- \frac{ V_s(b_{1;j}) - k_a |\uu_{1;j}\rb_{1;j}|\: }{T} \Big] 
\end{equation}
over the Boltzmann factors for all $n_1$ potential candidates $j$.
We continue in the same spirit and construct a new trial chain monomer
by monomer: Given the ($\alpha-1$)th monomer, we identify all $n_{\alpha}$ 
solvent particles at the distance $b\in[b_0 - b_s,b_0 + b_s]$ from the 
$(\alpha-1)$th monomer, and pick one with the probability 
\begin{eqnarray}
\lefteqn{P_{\alpha} = \frac{1}{Z_{\alpha}}
\exp\Big\{- \frac{1}{T} \big[\: V_s(b_{\alpha}) 
}\nonumber\\ \quad
&&   -k_a( |\uu_{\alpha-1} \rb_{\alpha}| 
              +  2 \rb_{\alpha-1} \rb_{\alpha} 
              + |\uu_{\alpha} \rb_{\alpha}|)  \big]  \Big\} 
\\
\lefteqn{\mbox{with}} \quad \nonumber \\
\lefteqn{Z_{\alpha} = \sum_{j = 1}^{n_{\alpha}} 
       \exp\Big\{- \frac{1}{T} \big[ \: V_s(b_{\alpha;j}) 
}\nonumber\\ \quad
&&  -k_a ( |\uu_{\alpha-1} \rb_{\alpha;j}| + 
                \rb_{\alpha-1} \rb_{\alpha;j} +
                |\uu_{\alpha;j} \rb_{\alpha;j}|)  \big] \Big\} .
\end{eqnarray}
The total probability of constructing a given new chain with $M$ monomers is
\begin{equation}
 P_{\mbox{\tiny new}} = \prod_{\alpha = 1}^{M} P_{\alpha}.
\end{equation}
For the final Metropolis step, one also has to compute the probability 
$P_{\mbox{\tiny old}}$ that the original, initially removed chain is
constructed. The way to do this is exactly analogous, except that 
the monomers are not picked randomly. The new chain is accepted with the 
probability
\begin{equation}
W_{\mbox{\tiny old} \to \mbox{\tiny new}} =
\min \big( 1, \frac{P_{\mbox{\tiny old}}}{P_{\mbox{\tiny new}}}
\exp (- \Delta E/T) \big),
\end{equation}
where $\Delta E$ is the energy difference between the old and the new
chain. Since no particles were moved, $\Delta E$ only contains
contributions of the bond energies. If the move is not accepted, the
original chain is restored.

A similar method has been described recently by Wijmans et al~\cite{chris}.
The implementation of this algorithm was essential for the success of
our simulations. With the new algorithm, the configurations equilibrated 
within a million Monte Carlo steps and reliable data for a range of 
grafting densities could be collected.

\section{Results}
\label{sec:results}

\begin{figure}[t]
\noindent
\fig{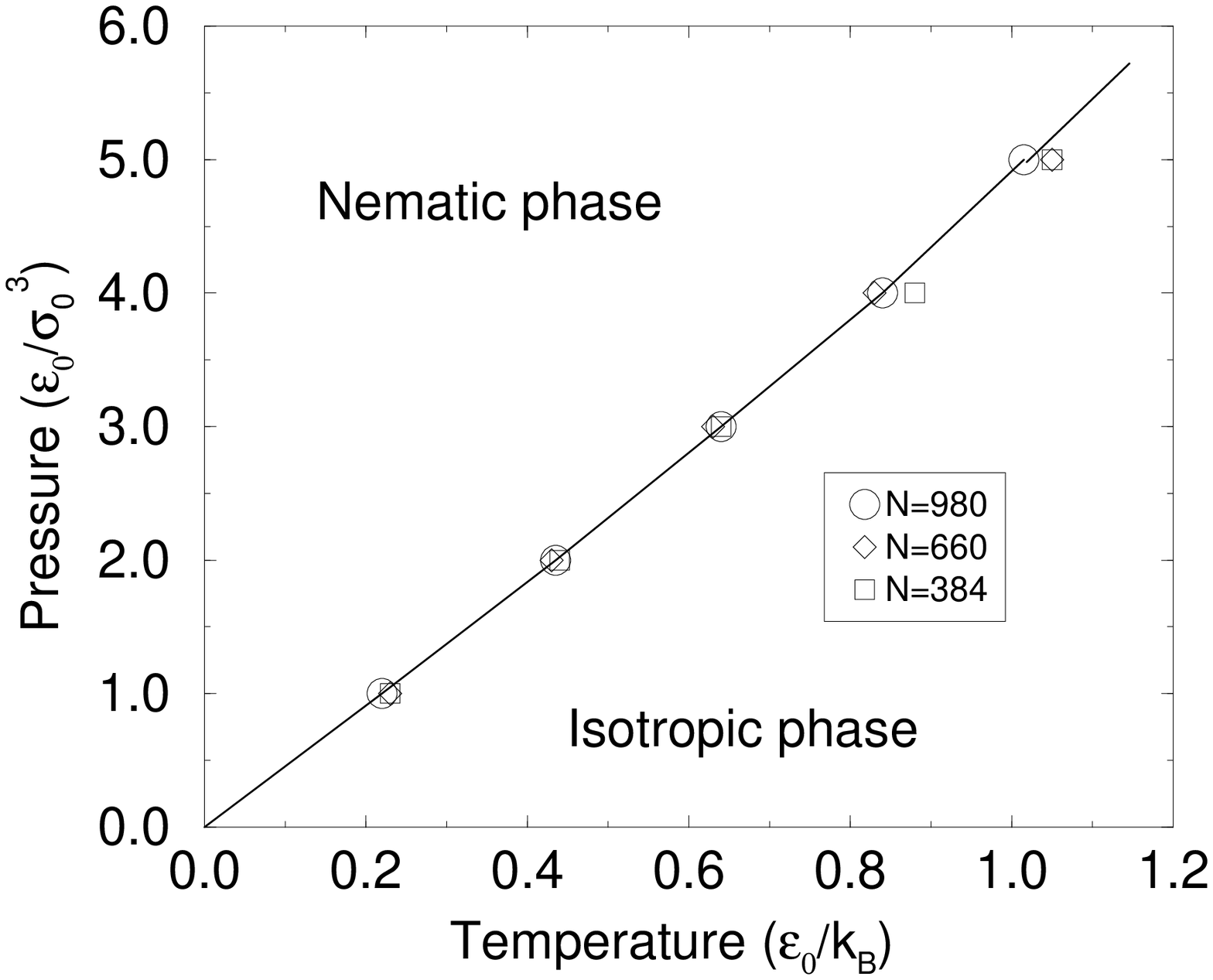}{60}{53}{}
\CCphbulk
\end{figure}

\begin{figure}[t]
\noindent
\fig{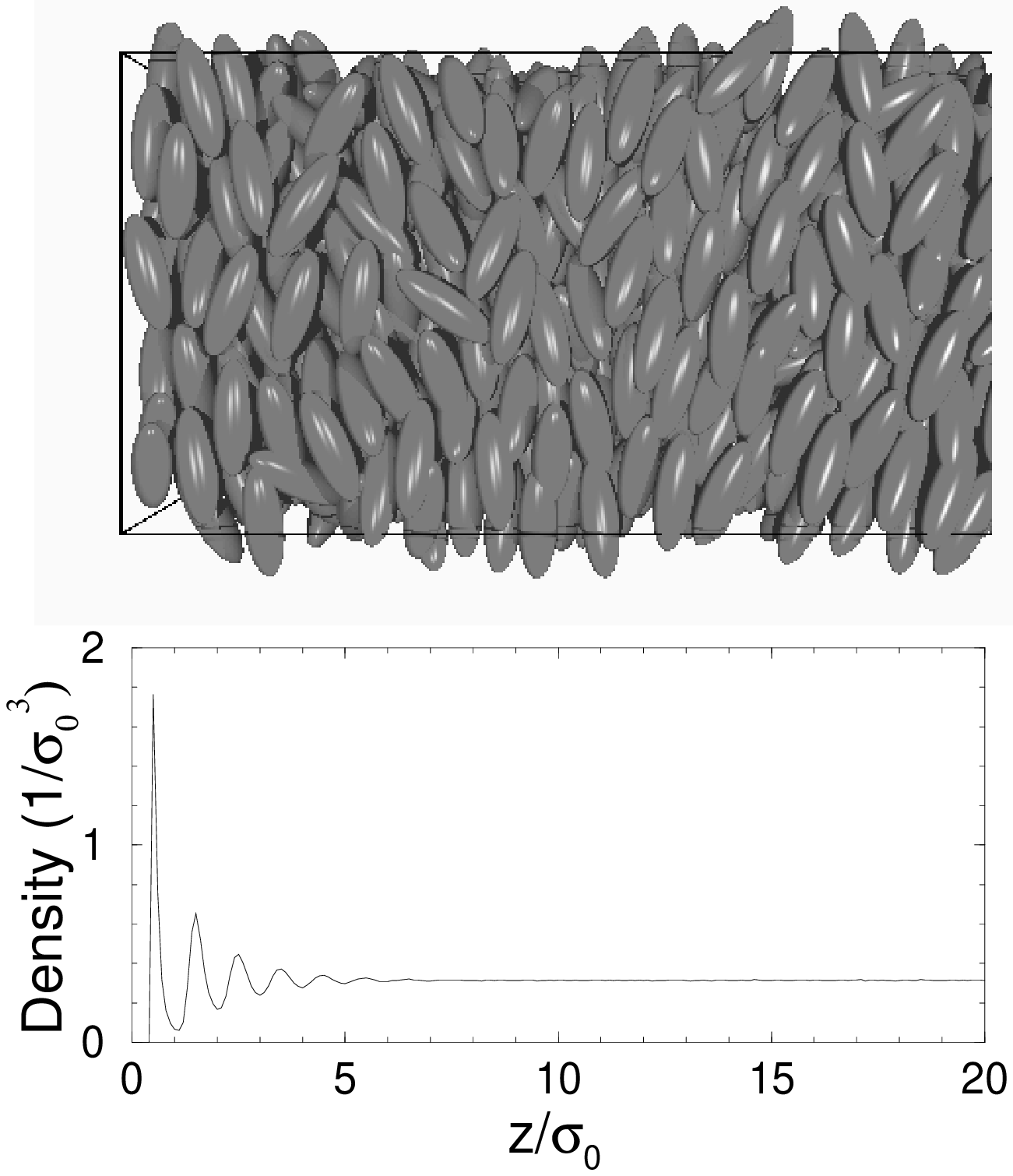}{60}{70}{}
\CCbare
\end{figure}

Before investigating the full system, we have characterized the 
different constituents of our system separately. 

The bulk phase diagram of the solvent particles was determined
from simulations of cubic systems with periodic boundary conditions 
in all directions. The resulting phase diagram in $P-T$ space is shown 
in Figure \ref{fig:phbulk}. The state point of our later simulations 
at $T=0.5$, $P=3$ is well in the nematic phase. The coexistence densities
at $P=3$ are $\rho_{\mbox{\tiny Nematic}} \approx 0.29$, and
$\rho_{\mbox{\tiny Isotropic}} \approx 0.28$. The systems were too small 
to allow for an accurate determination of the coexistence gap in
$\rho-T$ space. In $P-T$ space, however, the phase boundaries are not
affected much by finite size effects (Figure \ref{fig:phbulk}).

A bare substrate without grafted chains orients the particles
parallel to itself. A snapshot is shown in Figure \ref{fig:bare}.
The density profile reveals strong layering effects in 
close vicinity to the wall, which decay rapidly in the bulk.

\begin{figure}[t]
\noindent
\centerline{\fig{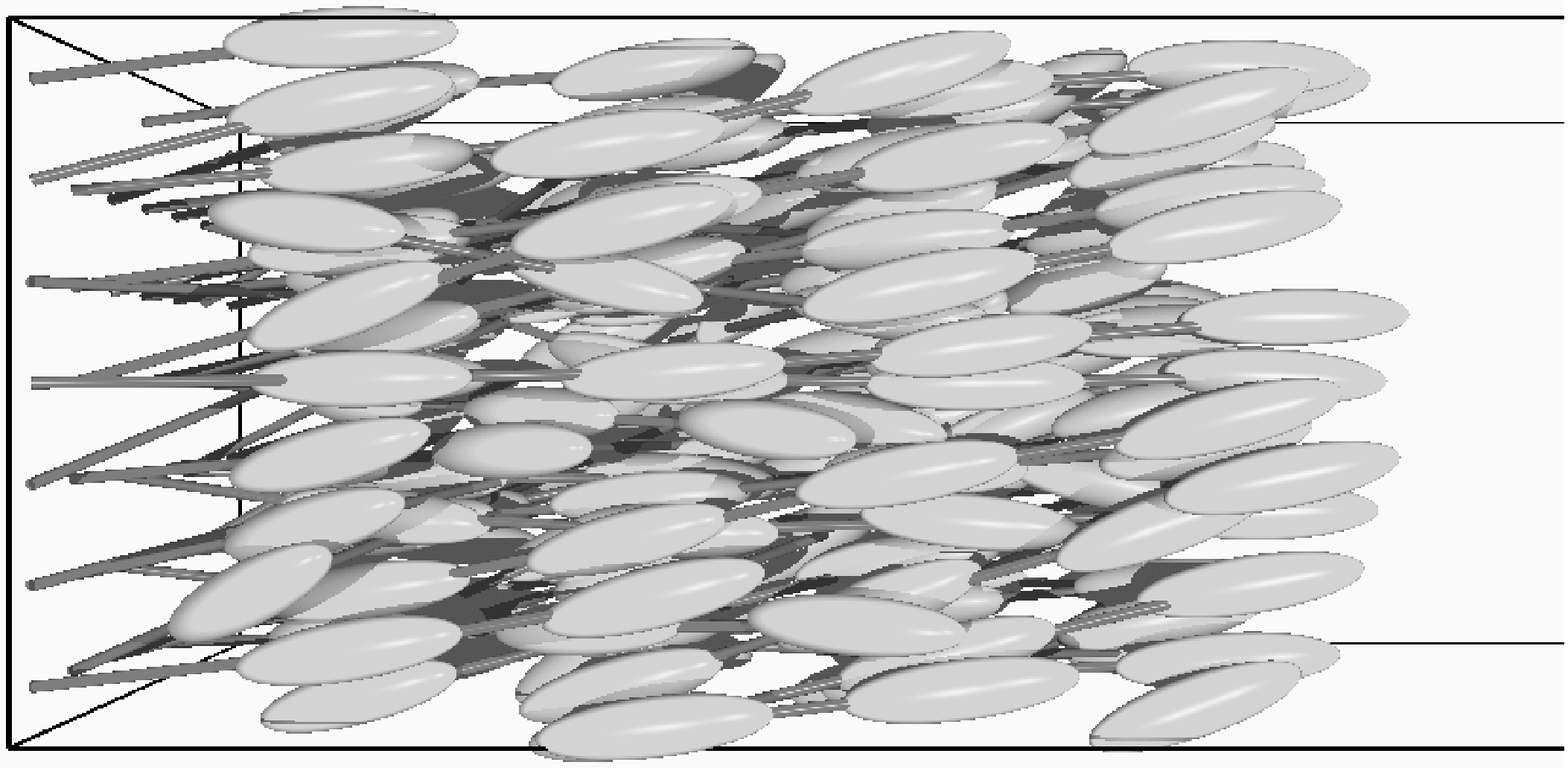}{60}{35}{}}
\CCsnapns
\end{figure}

\begin{figure}[t]
\noindent
\fig{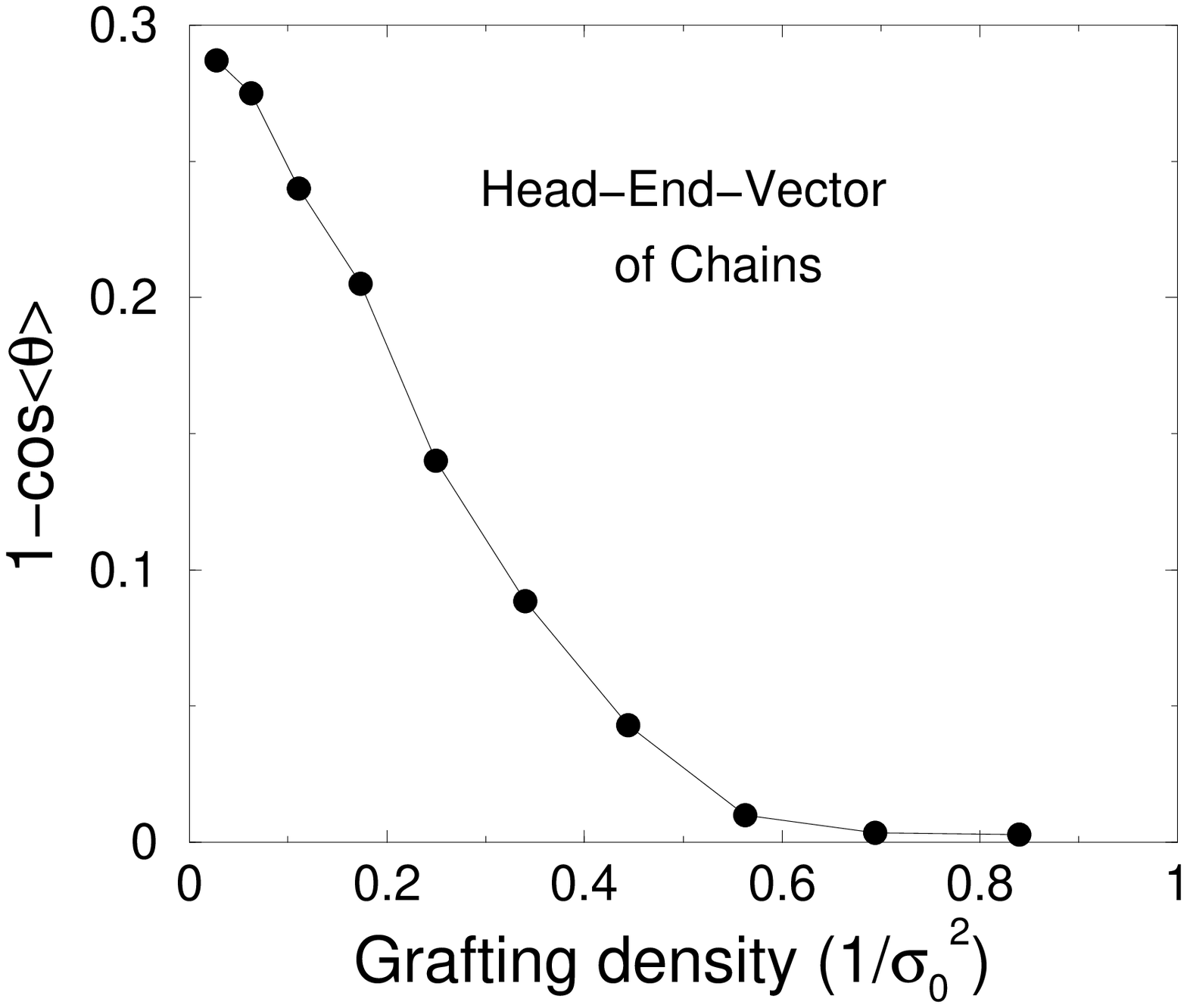}{60}{50}{}
\CCnosolv
\end{figure}

Finally, we have studied the behavior of brushes without solvent. 
A snapshot for a system with the moderate grafting density 
$\Sigma=0.34/\sigma_0^2$ is shown in Figure \ref{fig:snap_nosolv}.
The chains are disordered, with chain orientations distributed around 
$\theta=0$ ($\theta$ being the angle with respect to the substrate normal). 
The average angle $\langle \theta \rangle$ of the head-to-end vectors 
decreases rapidly with increasing grafting density $\Sigma$
(Figure \ref{fig:nosolv}).

\begin{figure}[b]
\noindent
\centerline{\fig{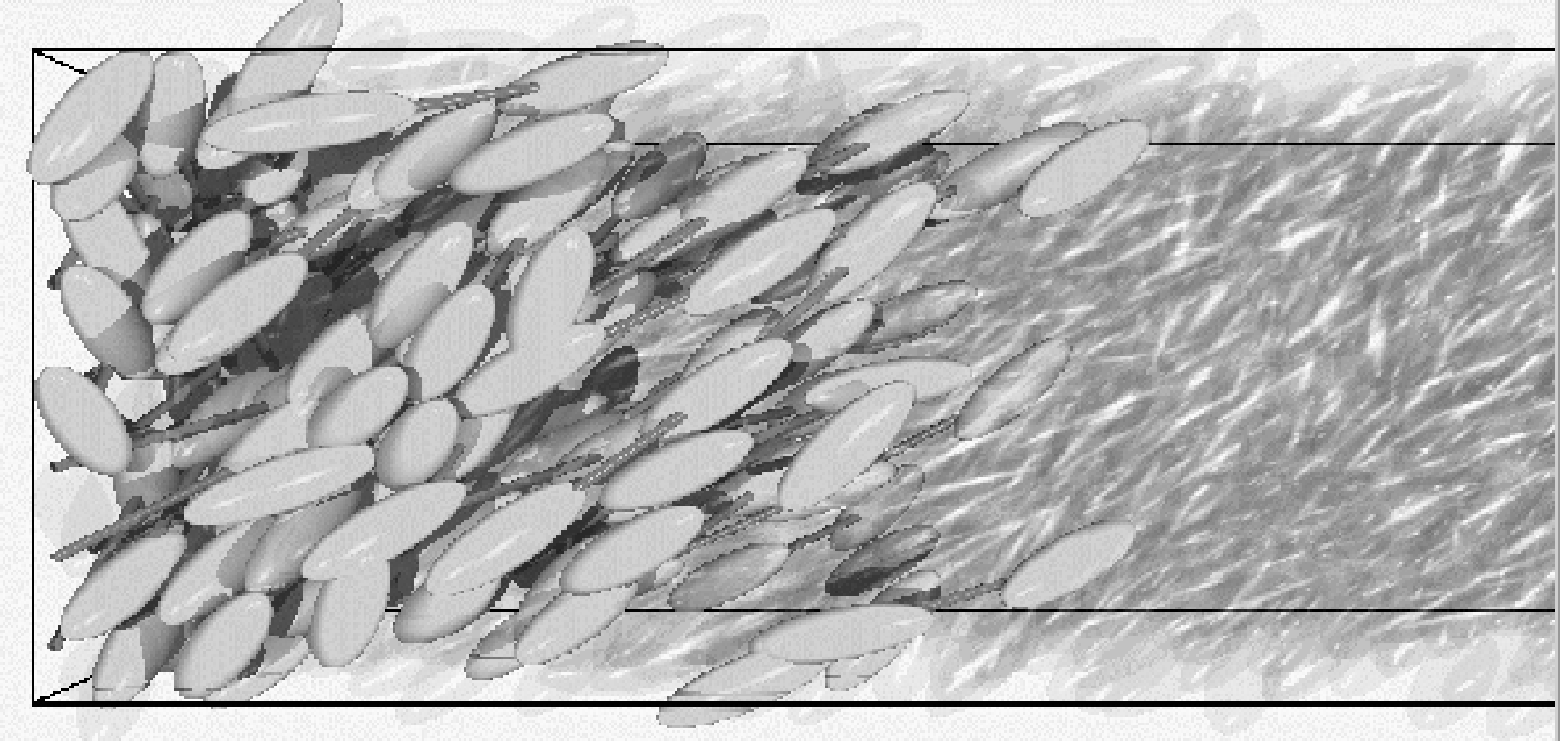}{70}{40}{}}
\centerline{\fig{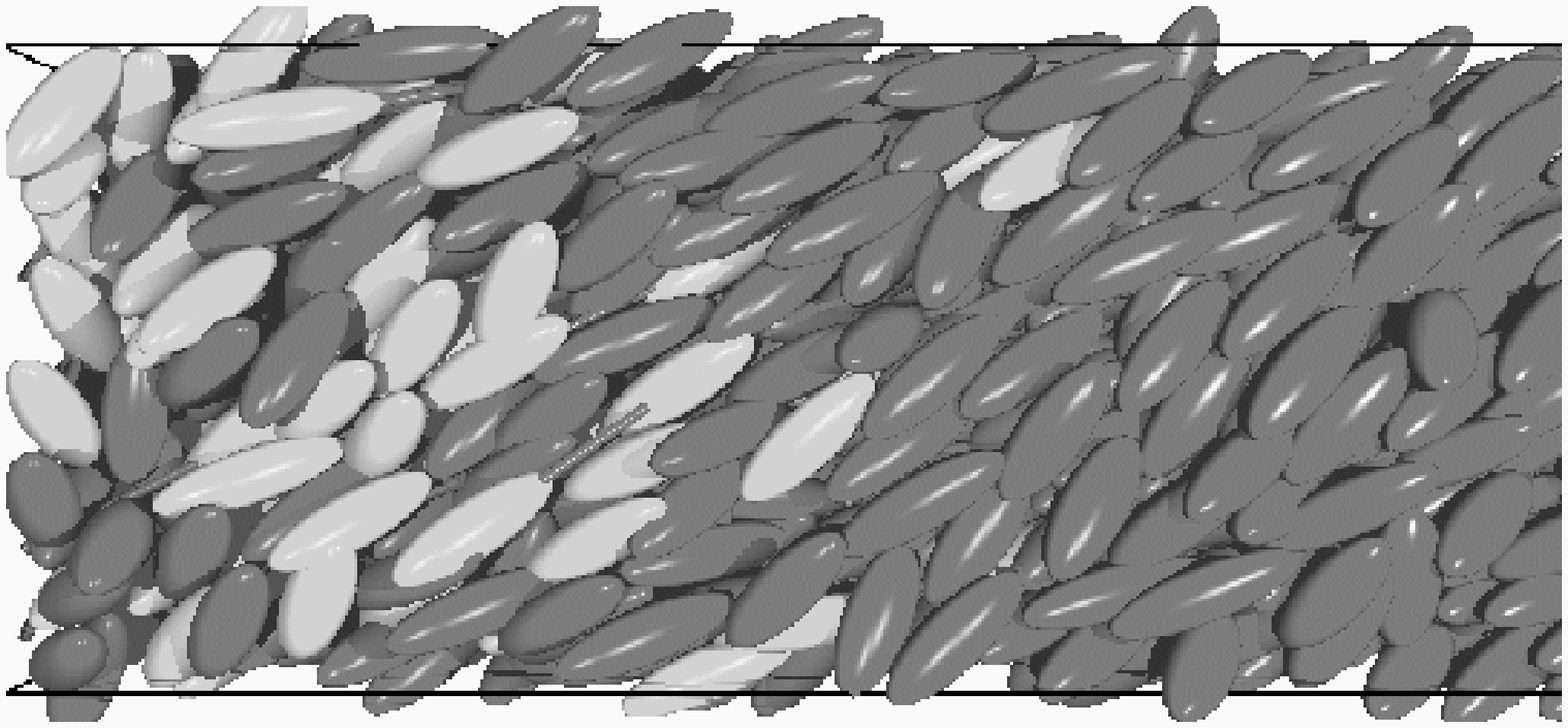}{70}{40}{}}
\CCsnap
\end{figure}

We turn to the discussion of the results for the full system.
Figure \ref{fig:snap} shows a snapshot of a brush with solvent at the 
grafting density $\Sigma=0.34/\sigma_0^2$. Comparing this figure with
Figure \ref{fig:nosolv}, we note that the presence of the solvent makes
the brush much more compact, i.e., the solvent particles introduce 
an effective attractive interaction between the chains. The chains are
tilted in a common direction (Figure \ref{fig:snap} (top)). 
The tilt propagates into the bulk of the system outside of the brush 
(Figure \ref{fig:snap} (bottom)). As anticipated in the
introduction, we thus find that a brush can induce tilt in the 
orientation of a nematic solvent.

The comparison of Figure \ref{fig:nosolv} and Figure \ref{fig:theta} 
(top) shows that the average tilt angle $\langle \theta \rangle$ of 
the head-to-end vector of chains increases in the presence of the 
solvent at all grafting densities. Most notably, the chains remained
tilted up to the highest grafting densities that we considered. 
However, the tilt of the chains does not always produce tilted anchoring 
in the nematic bulk fluid.  Figure \ref{fig:theta} (bottom) demonstrates 
that the solvent particles have less tilt than the chain particles, and 
stand almost perpendicular to the surface at the highest grafting density.
In fact, a closer inspection of our data shows that the remaining
average angle $\langle \theta \rangle$ stems mainly from the
contribution of the solvent particles inside of the brush, and
from orientational disorder of the particles outside of the brush.
A more detailed discussion will be presented elsewhere~\cite{harald3}.

Figure~\ref{fig:theta} (bottom) reveals a second feature of
the system: At low grafting densities, the wall aligns the solvent 
particles in a planar way, i. e., the alignment is basically 
determined by the anchoring properties of the bare substrate.
At the grafting density $\Sigma \approx 0.12$, the tilt angle of 
solvent particles jumps discontinuously to a lower value. 
The data thus indicate that there is  a first order transition 
between planar and tilted anchoring.

\noindent
\begin{figure}[b]
\fig{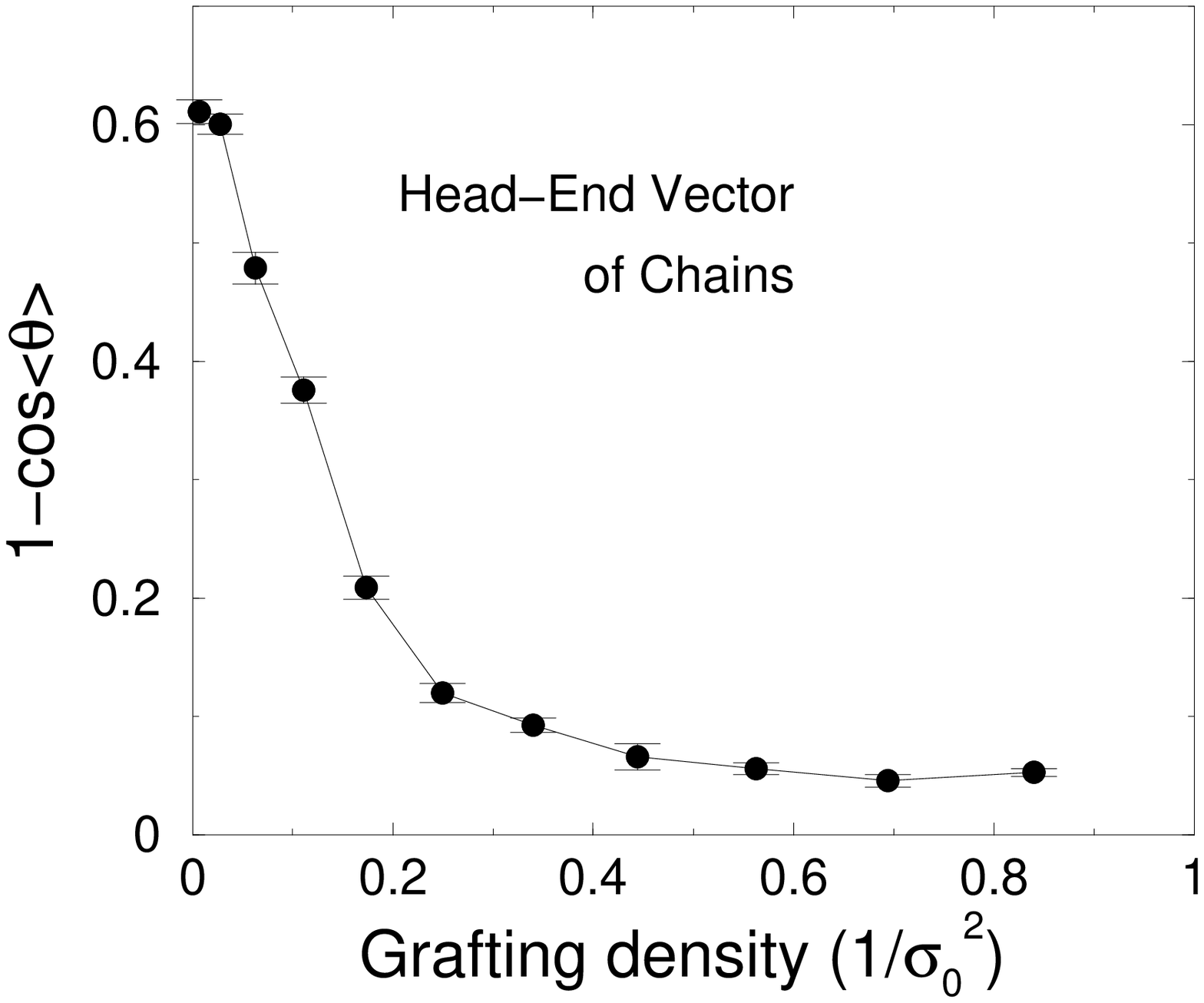}{60}{50}{} 
\fig{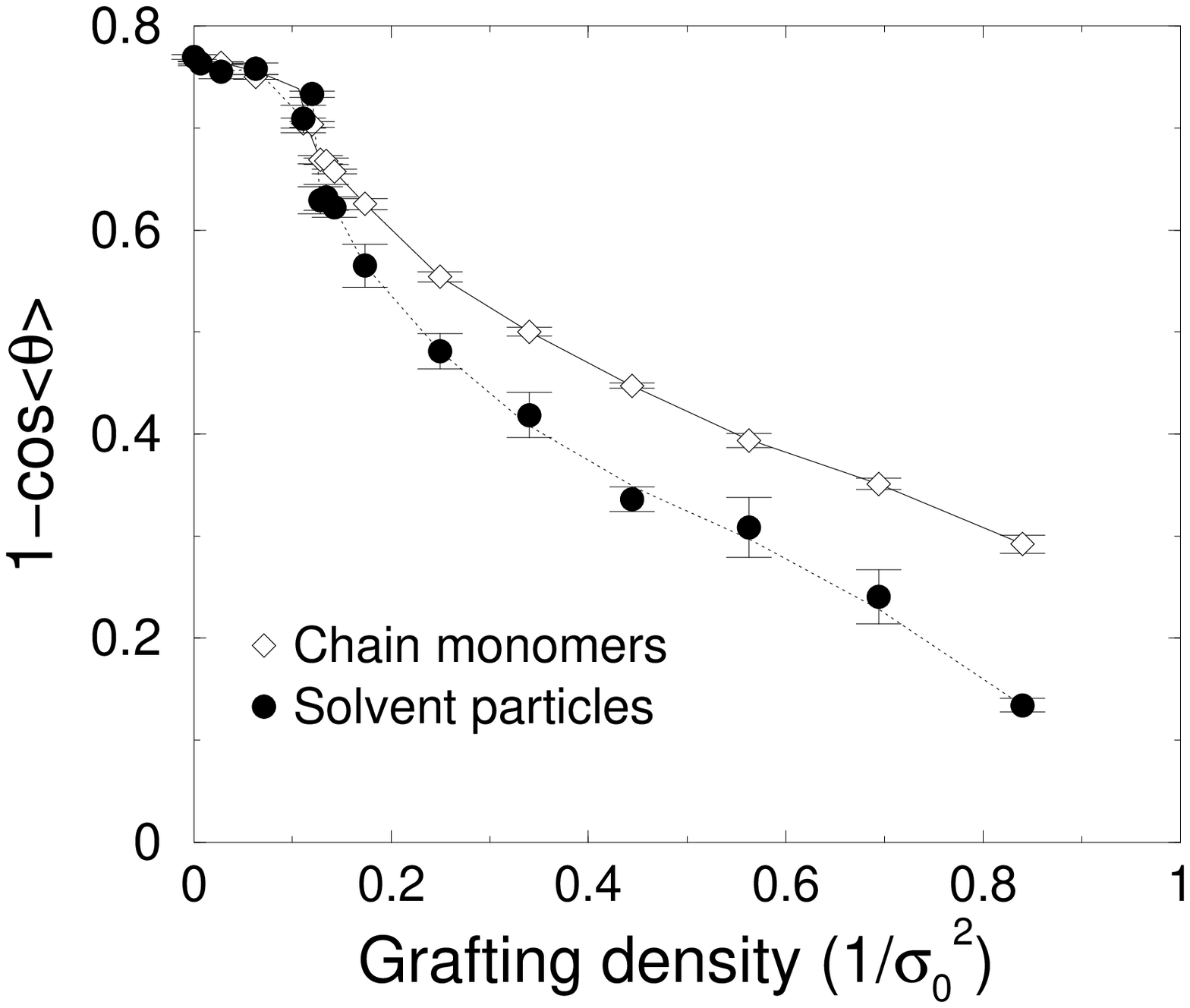}{60}{50}{}
\CCtheta
\end{figure}

\section{Discussion and Summary}

Our simulation data disclose the following scenario: 
We find three different anchoring regimes. At low grafting densities, 
the anchoring on the wall is planar. At high grafting densities, 
it is homeotropic, i.e., perpendicular to the wall. Tilted anchoring 
can be achieved in a range of intermediate grafting densities. 
The transition between planar and tilted anchoring is presumably 
first order. The transition between tilted and perpendicular anchoring
has not been studied in detail, because simulations at high grafting
densities are very expensive. The data are consistent with
theoretical mean field considerations~\cite{harald1}, which
suggest that it is continuous. According to theory, the anchoring
behavior at high grafting densities is controlled by the structure
of the interface between the polymer layer and the nematic fluid. 
The competition between attractive chain interactions, the
translational entropy of the solvent, and its elasticity, drives
a continuous transition from tilted to homeotropic alignment.

From a practical point of view, low grafting densities are more
interesting, because they can be realized more easily in experiments. 
The anchoring transition which we observe in that regime resembles that 
predicted by Halperin and Williams~\cite{avi1}. However, the underlying 
mechanism is probably quite different. The theory of Halperin and 
Williams relies on the existence of hairpins, abrupt reversals in the 
directions of the chains, whereas our ``polymers'' are much too short 
to support such defects. Nevertheless, the transition seems to be
controlled by the organization of the chains in the polymer layer.
We shall investigate this in more detail in a future 
publication~\cite{harald3}.

\section*{Acknowledgments}

We thank M.~P.~Allen for invaluable discussions. Furthermore, we have
benefitted from useful conversations with K.~Binder, D.~Johannsmann, 
J.~R\"uhe, and A.~Halperin. M.~P.~Allen and K.~Binder were so
kind to let us perform a major part of the simulations on the
computers of their groups. This work was funded by the German
Science Foundation (DFG).

\end{document}